\documentclass{csmagclass}
\usepackage[utf8]{inputenc}
\usepackage{multicol}
\usepackage[numbers,square,sort&compress,comma]{natbib}

\usepackage{epstopdf}
\usepackage{adjustbox}
\usepackage{graphicx}
\PassOptionsToPackage{monochrome}{xcolor}
\graphicspath{{img-gray/}}
\usepackage[refpage]{nomencl}
\usepackage{longtable}
\usepackage{caption}
\usepackage{subcaption}
\usepackage{float}
\usepackage{multirow}
\usepackage{array}
\usepackage{amsmath}
\usepackage{amsfonts}
\usepackage{pdfpages}
\usepackage{amsmath}
\usepackage{amsfonts}
\usepackage{txfonts}

\title{Magnetocaloric properties of an Ising antiferromagnet on a kagome lattice}
\author[1]{{ M. SEMJAN}}
\author[1]{{ M. ŽUKOVIČ\thanks{Corresponding author: milan.zukovic@upjs.sk}}}
\affil[1]{{Institute of Physics, Faculty of Science, 
P.J Šafárik University in Košice, Park Angelinum 9, 040 01 Košice, Slovakia}}

\newcommand{\myheight}{0.77\textwidth}
\newcommand{\mywidth}{0.46\textwidth}

\begin{document}

\maketitle

\begin{abstract}
Owing to a high degree of geometrical frustration an Ising antiferromagnet on a kagome lattice (IAKL) is known to exhibit no long-range ordering at any temperature, including the ground state. Nevertheless, at low temperatures it shows a strongly correlated, highly fluctuating regime known as a cooperative paramagnet or classical spin liquid. In the ground state it is characterized by a macroscopic degeneracy which translates to a relatively large value of the residual entropy. It has been shown that the presence of a macroscopic degeneracy associated with geometrical frustration below the saturation field can facilitate an enhanced magnetocaloric effect (MCE), which can exceed that of an ideal paramagnet with equivalent spin by more than an order of magnitude. In the present study we investigate magnetic and magnetocaloric properties of IAKL by Monte Carlo simulation. In particular, we calculate the entropy of the system using the thermodynamic integration method and evaluate quantities which characterize MCE, such as the isothermal entropy and adiabatic temperature changes in a varying magnetic field. It is found that IAKL shows the most interesting magnetocaloric properties at low temperatures and moderate magnetic fields, suggesting its potential to be used in technological applications for low-temperature magnetic refrigeration. 
\end{abstract}

%\begin{multicols}{2}
\section{Introduction}
\begin{figure}
    \center
    \includegraphics[width = 0.3\textwidth]{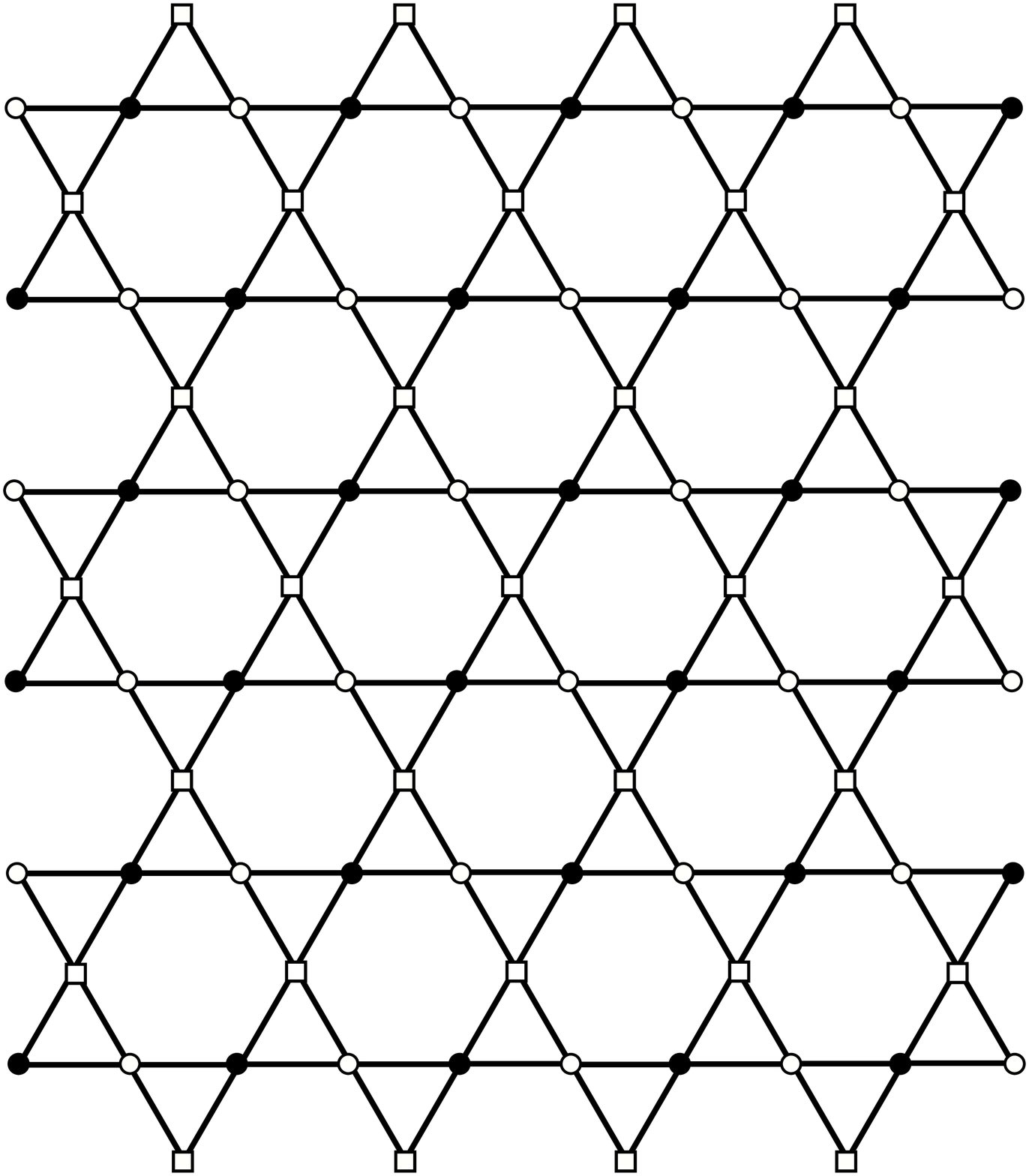}
    \caption{The kagome lattice is a tripartite lattice and can be divided into three sublattices - denoted by $\medcirc$, $\medbullet$ and 
    $\square$.}
    \label{fig:lattice}
\end{figure}

\paragraph{}
The phenomenon of geometrical frustration in magnetic systems is closely related to the geometry of the lattice, which does not allow to fully satisfy all the interactions between its spins \cite{diep2013}. The effects of frustration are rich and still not well-understood. Previous research suggests that the field-induced adiabatic temperature change is significantly larger for such systems \cite{zhitomirsky2003} than for their non-frustrated counterparts, which makes them better candidates for magnetic refrigeration using the magnetocaloric effect (MCE). MCE can by characterized be the change of the magnetic entropy in response to variation of the magnetic field. The Ising antiferromagnet on a kagome lattice (IAKL) is a great example of a highly frustrated system, which was extensively studied in the past \cite{diep2013, syozi1951, kano1953,  shores2005}. The kagome lattice consists of corner-sharing triangles (Fig. \ref{fig:lattice}) and it's elementary cell is shaped like the `Star of David`. The exact solution for the Ising model on the kagome lattice was found in 1951 by I. Syozi \cite{syozi1951}. He discovered that in the ferromagnetic case a specific heat diverges at the temperature $3+2\sqrt{3}$, which is higher than the critical temperature of the square lattice ($3+2\sqrt{2}$). Nevertheless, the antiferromagnetic case shows no critical behavior at any temperature. It is also known that the density of a residual entropy of IAKL is 0.5018$k_B$ \cite{kano1953}, which is larger than that of the triangular lattice (0.3231$k_B$) \cite{wannier1950}. The ground state energy per spin was calculated by Kano and Naya \cite{kano1953} and it is  $-2J$ in the ferromagnetic case and $-2J/3$ in the antiferromagnetic case. In this paper, magnetocaloric properties of IAKL in the presence of an external magnetic field are explored by means of Monte Carlo simulation. 

\section{Model}
\paragraph{}
The Hamiltonian of the studied system is given by
\begin{equation}
 \mathcal{H} = -J\sum_{\langle i,j\rangle }{\sigma_i\sigma_j} -h\sum_{i=1}^N{\sigma_i}, \label{eq:H}
\end{equation}
where the first summation goes over the nearest neighbors, the second sum goes over each spin, $N$ is the total number of spins, $\sigma_i = \pm 1$ and $h$ is the external magnetic field. In order to introduce frustration, interactions between neighboring spins were chosen to be antiferromagnetic ($J<0$).

\section{Method}
\paragraph{}
The standard Metropolis algorithm was used. At each Monte Carlo (MC) step a new state $\sigma'_i$ is proposed for a selected spin $\sigma_i$ and the new configuration is accepted with the probability $p(\sigma_i \rightarrow \sigma'_i) = \min\{1,\exp(-\beta dE)\}$, where $\beta = 1/(k_BT)$ is the inverse temperature and $dE$ is the energy difference between the proposed and the old configuration. When the algorithm makes a MC trial for each spin, we say that a MC sweep was completed. From the simulation, we directly obtain the energy per spin $e = \langle\mathcal{H} \rangle/(|J|N)$,  from Eq. (\ref{eq:H}), and the magnetization per spin $m = \langle M \rangle/N$, where $M = \sum_{i=1}^N{\sigma_i}$ and $\langle ... \rangle$ denotes a thermal average. However, the entropy, which we are interested in, can not be calculated directly from the MC simulation. Nevertheless, it can be obtained as a function of the inverse temperature by utilizing the Thermodynamic Integration Method (TIM) \cite{kirkpatrick1977} as:
\begin{equation}
S(\beta) = N\ln{(2s+1)} + \beta E(\beta) - \int_0^\beta E(\beta')d\beta',\label{eq:ent} 
\end{equation}
where the spin number $s$ is in our case 1/2 and $E = N e$. MCE is characterized by the following quantities: the adiabatic temperature change $\Delta T_{ad}$ and the isothermal entropy change $\Delta S_{iso}$. For a fixed temperature $T$ and the change of the field from $h_1$ to $h_2$, $\Delta S_{iso}$ is defined as
\begin{equation}
    \Delta S_{iso}(h_2-h_1,T) = S(h_2,T) - S(h_1,T).
    \label{eq:ds}
\end{equation}
Similarly, in adiabatic condition with the entropy $S$ the corresponding temperature change $\Delta T_{ad}$ from $T_1$ to $T_2$ can be calculated as
\begin{equation}
    \Delta T_{ad}(h_2-h_1,S) = T_2(h_2) - T_1(h_1).
    \label{eq:dt}
\end{equation}

\begin{figure}[h]
    \begin{subfigure}[b]{\mywidth}
        \center
        \centering
        \includegraphics[height = \myheight, clip]{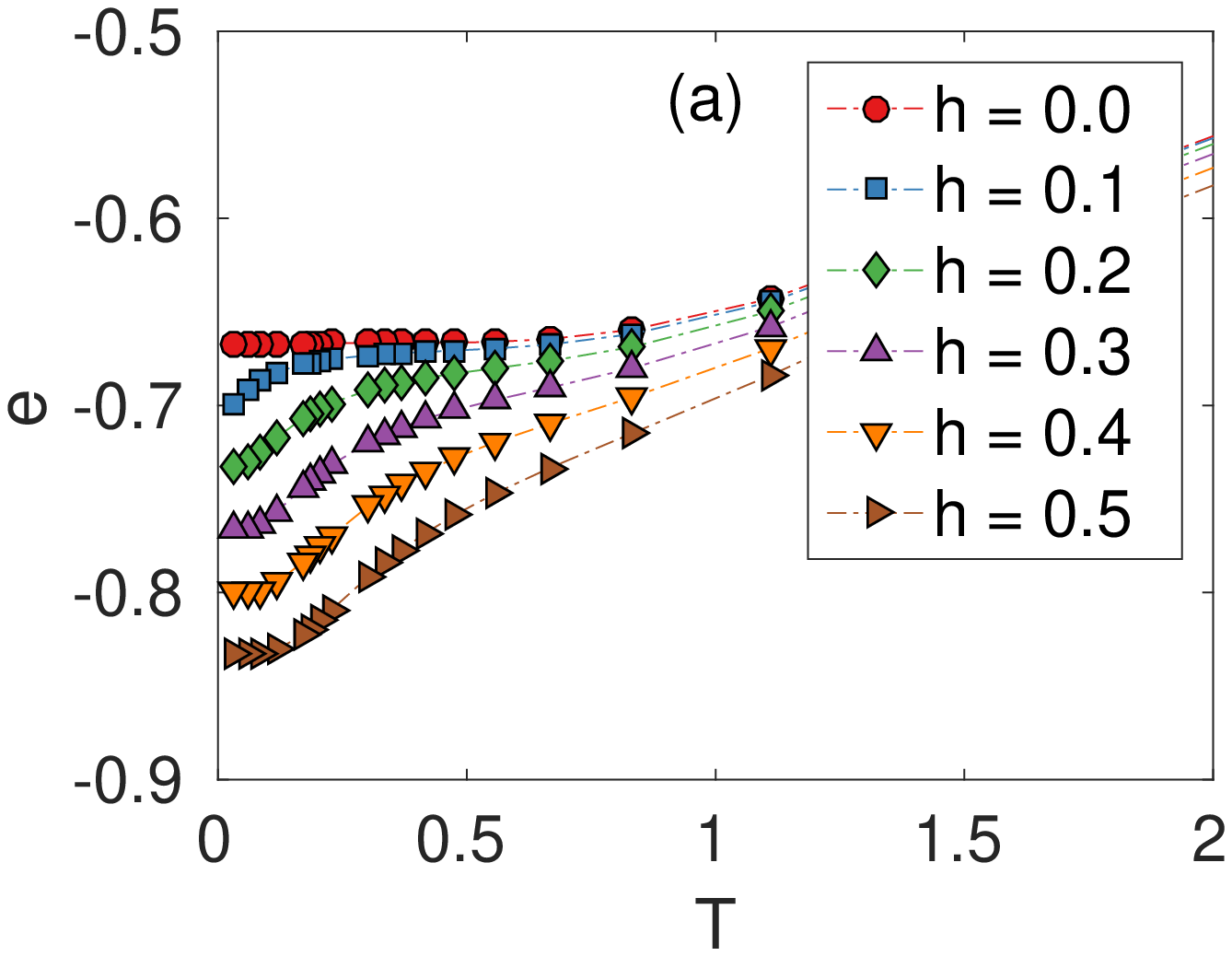}
        %\caption{}
        \label{fig:e1}
    \end{subfigure}
    \hfill
    \begin{subfigure}[b]{\mywidth}
        \center
        \centering
        \includegraphics[height = \myheight, clip]{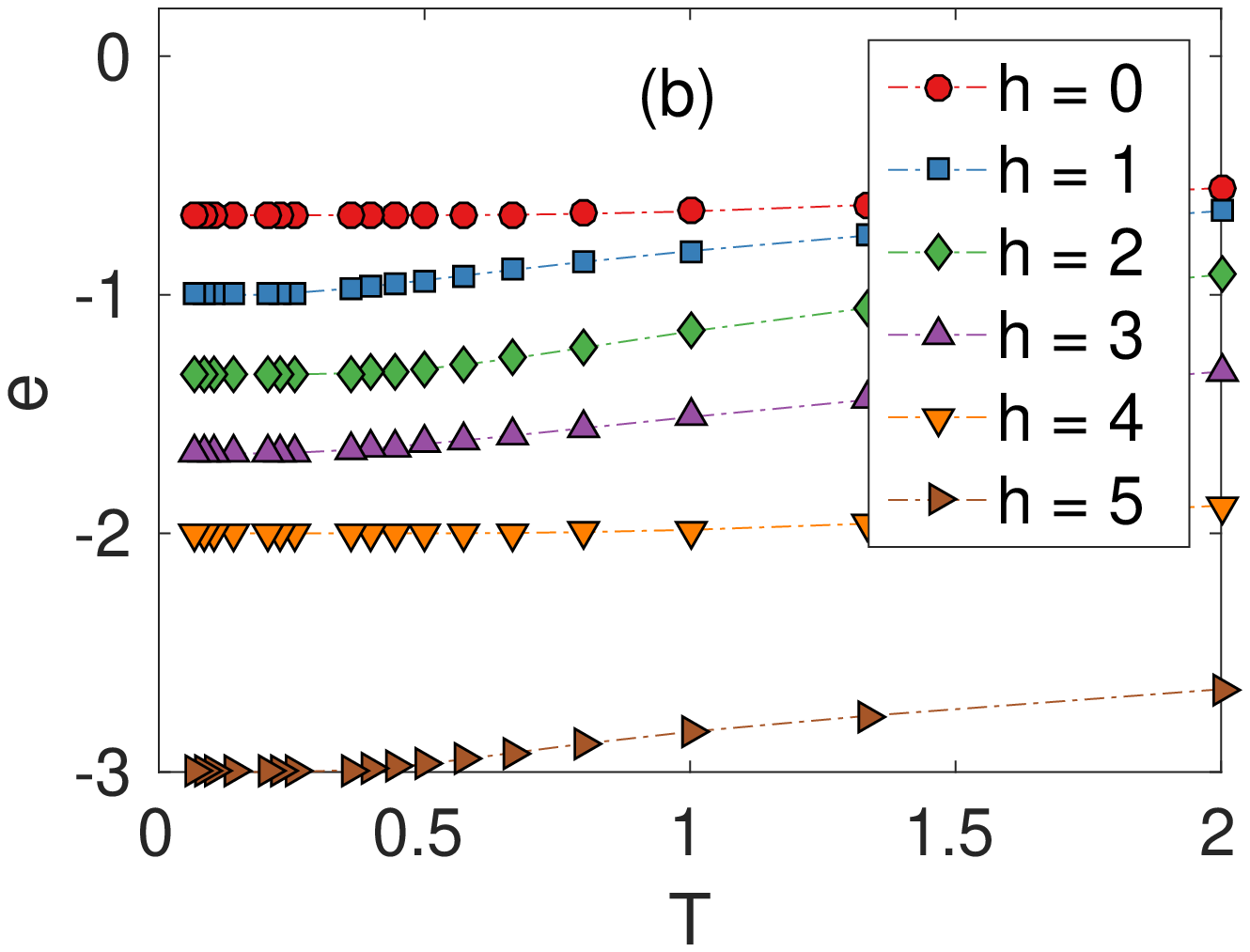}
        %\caption{}
        \label{fig:e2}
    \end{subfigure}
    \caption{Energy per spin for various values of the field $h$.}
    \label{fig:energie}
\end{figure}

\section{Results and discussion}
\paragraph{}
Throughout the paper we set $J = -1$ and $k_B = 1$. For each temperature 4$\times 10^5$ MC sweeps were used to calculate physical quantities after discarding another $10^5$ for thermalization. For smaller values of the field $h = 0$, $0.1$, $0.2$, $0.3$, $0.4$ and $0.5$, simulations were performed on the lattice with $50\times 50$ cells (7500 spins in total). The inverse temperature was chosen in the range $\beta\in\langle 0, 50 \rangle$. Values of $\beta$ were denser in a low-temperature region to increase the precision of TIM. Additional simulations were performed on a smaller lattice with $20\times 20$ cells (1200 spins in total) for larger values of the field $h = 1$, $2$, $3$, $4$ a $5$. In the absence of the field, our simulation yielded the ground state (GS) energy close to the value $e = -2/3$ (see Fig. \ref{fig:energie} (a)), which is in a good agreement with the exact value \cite{kano1953}. The energy corresponds to each elementary triangle having two spins up and one down or vice versa, with no ordering among them - the spin liquid state. In the presence of the external magnetic field, the GS energy is lowered by a Zeeman term proportionally to the field's strength. 
\iffalse
For $h=4$, one can observe that the fully polarized state is energetically more advantageous than spin liquid state and above this value of $h$ all spins point in the direction of the field (in Fig. \ref{fig:e2}). 
\fi
\begin{figure}[h]
    \begin{subfigure}[b]{\mywidth}
        \center
        \centering
        \includegraphics[height = \myheight, clip]{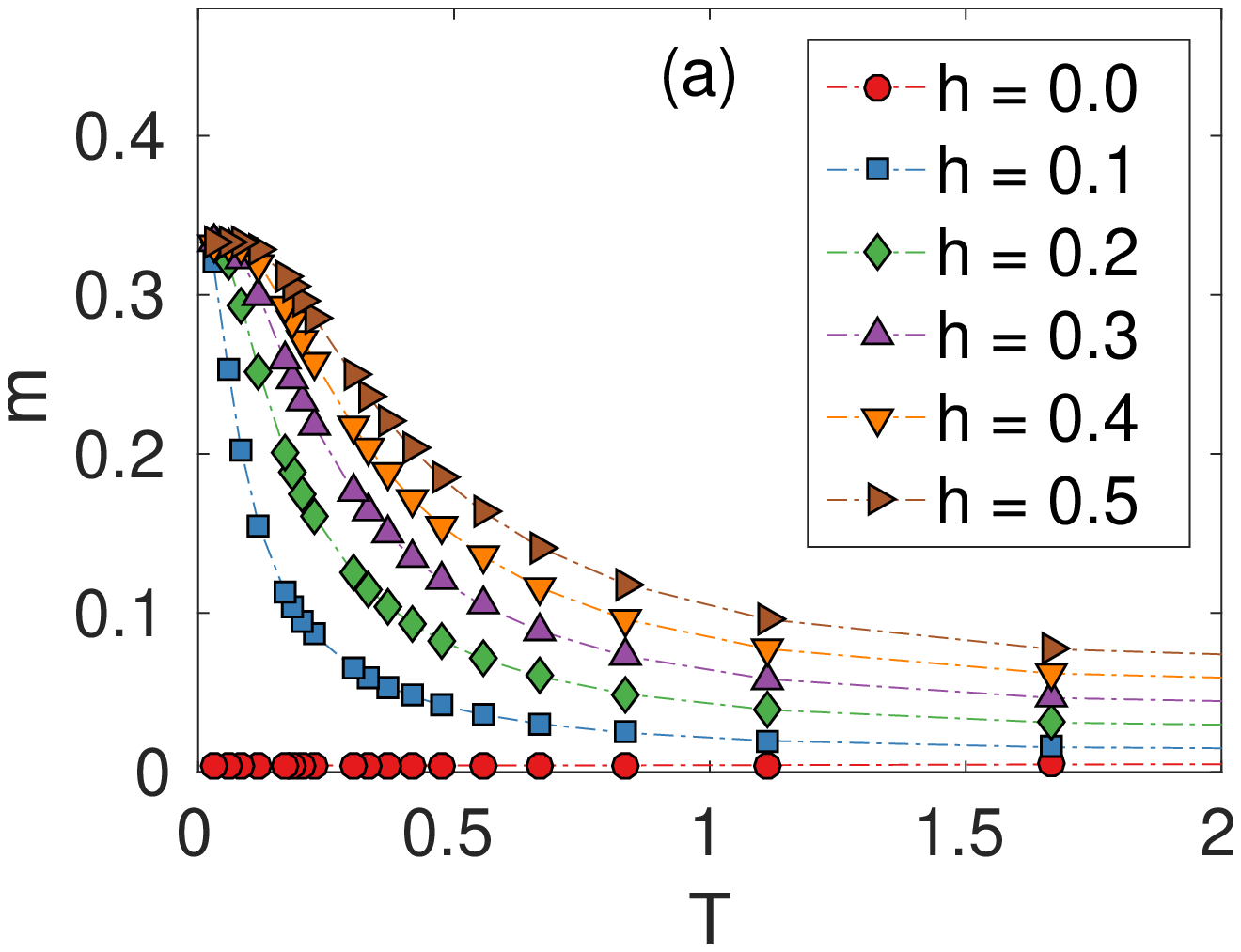}
        %\caption{}
        \label{fig:m1}
    \end{subfigure}
        \hfill
    \begin{subfigure}[b]{\mywidth}
        \center
        \centering
        \includegraphics[height = \myheight, clip]{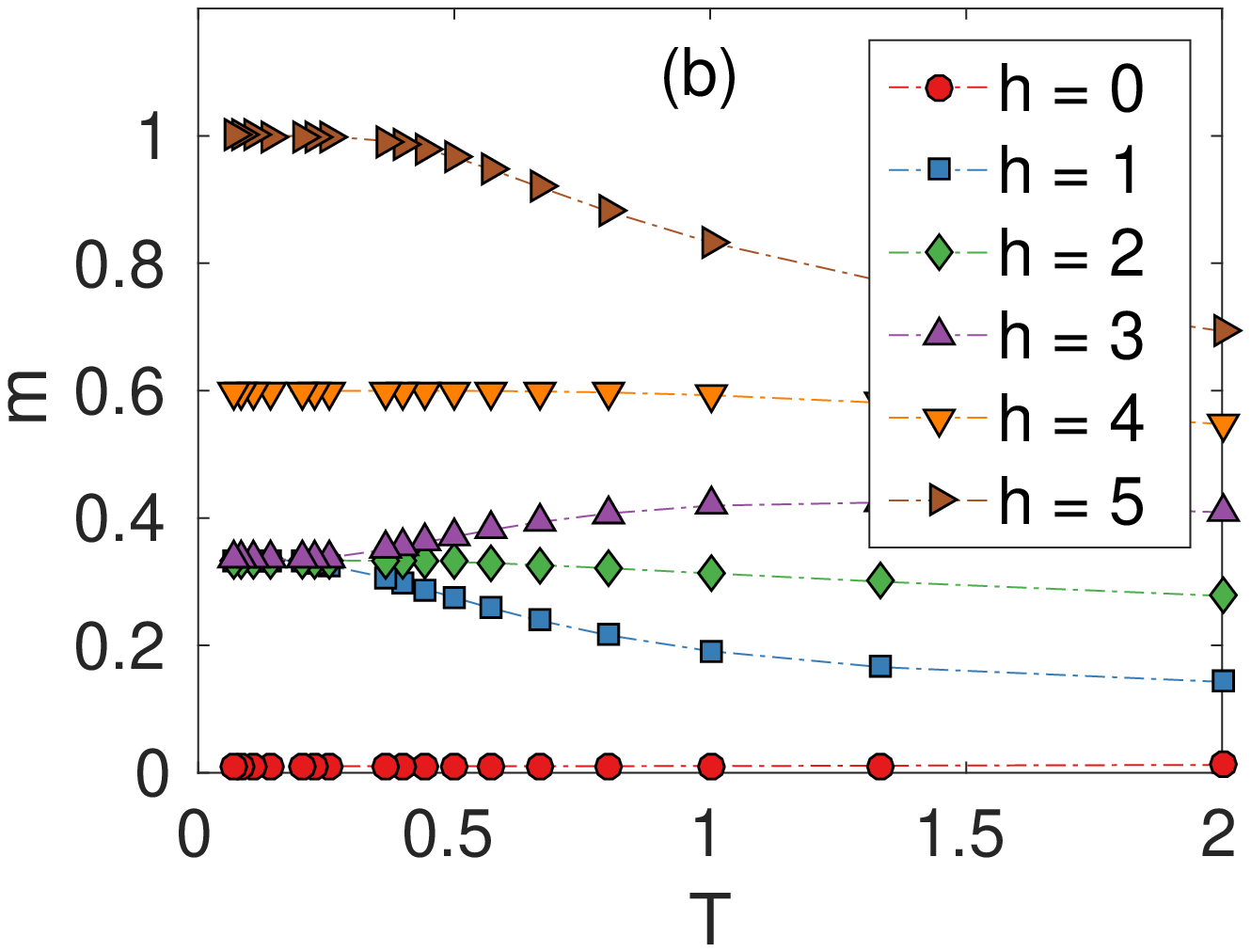}
        %\caption{}
        \label{fig:m2}
    \end{subfigure}
    \\
    \begin{subfigure}[b]{\mywidth}
        \center
        \centering
        \includegraphics[height = \myheight, clip]{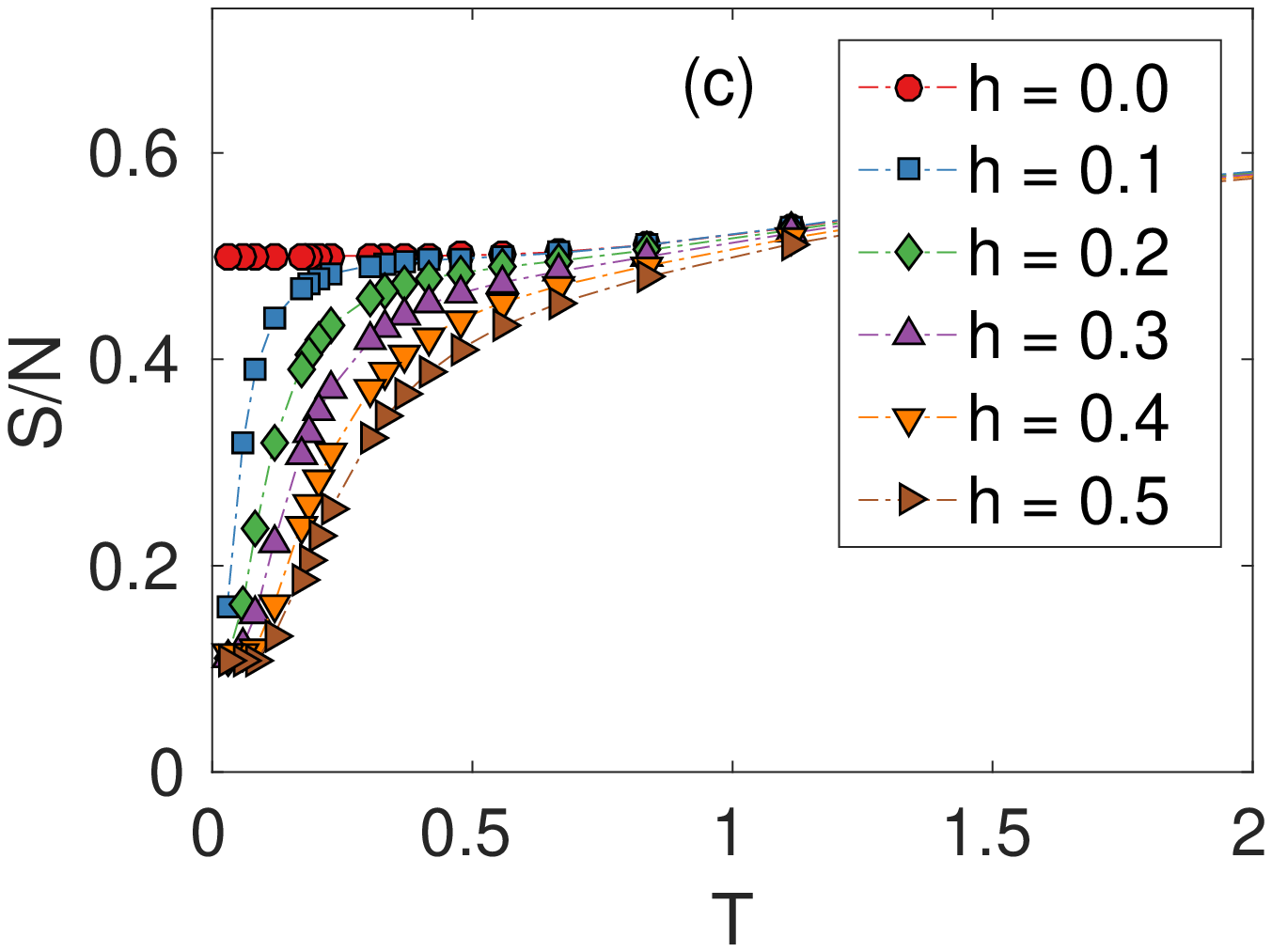}
        %\caption{}
        \label{fig:s1}
    \end{subfigure}
    \hfill
    \begin{subfigure}[b]{\mywidth}
        \center
        \centering
        \includegraphics[height = \myheight, clip]{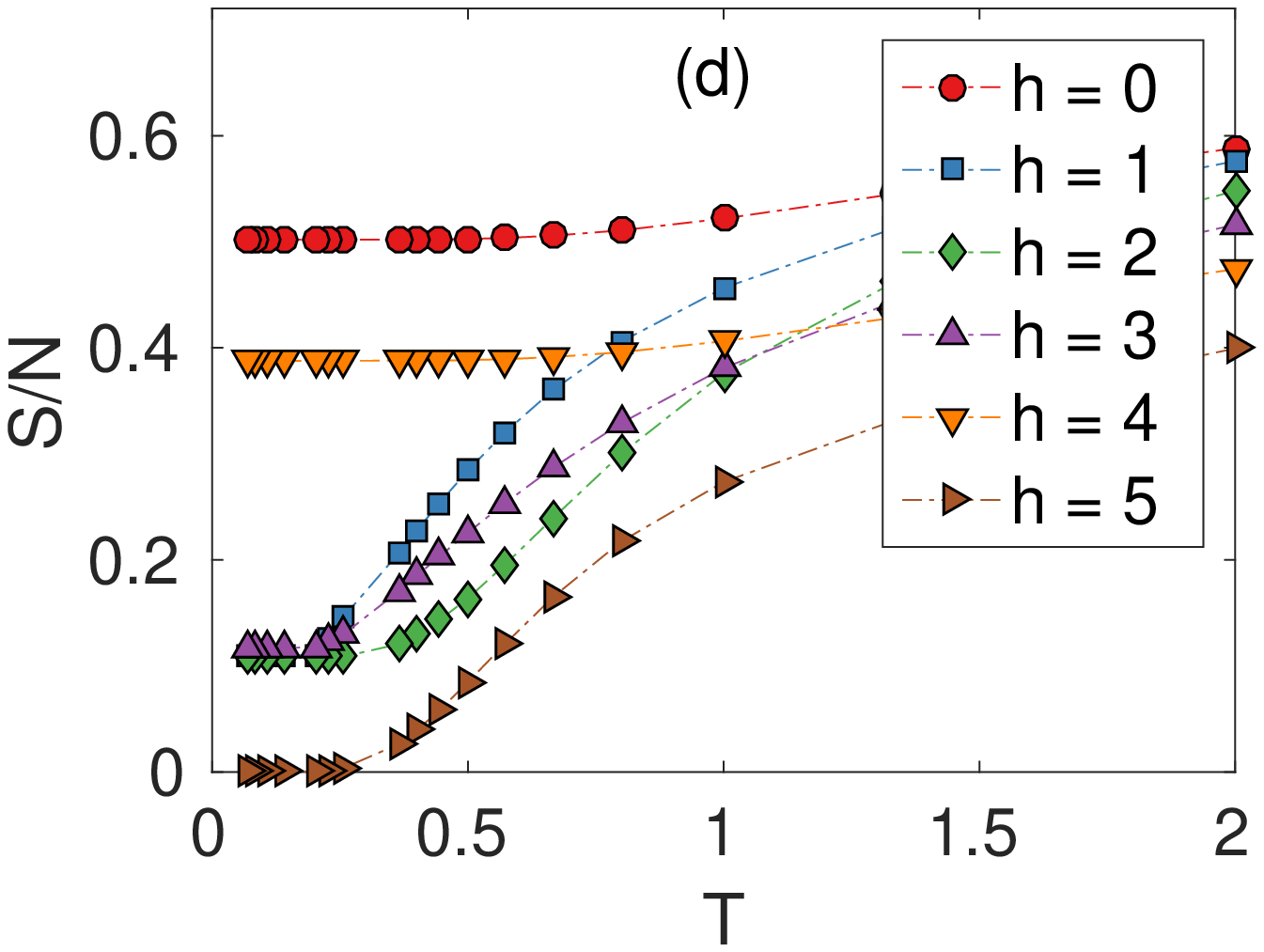}
        %\caption{}
        \label{fig:s2}
    \end{subfigure}
    \caption{Magnetization per spin and entropy density for various values of the field $h$.}
    \label{fig:mag}
\end{figure}

\begin{figure}[h]
    \centering
    \begin{subfigure}[b]{\mywidth}
    	\center
        \centering
        \includegraphics[height = \myheight, clip]{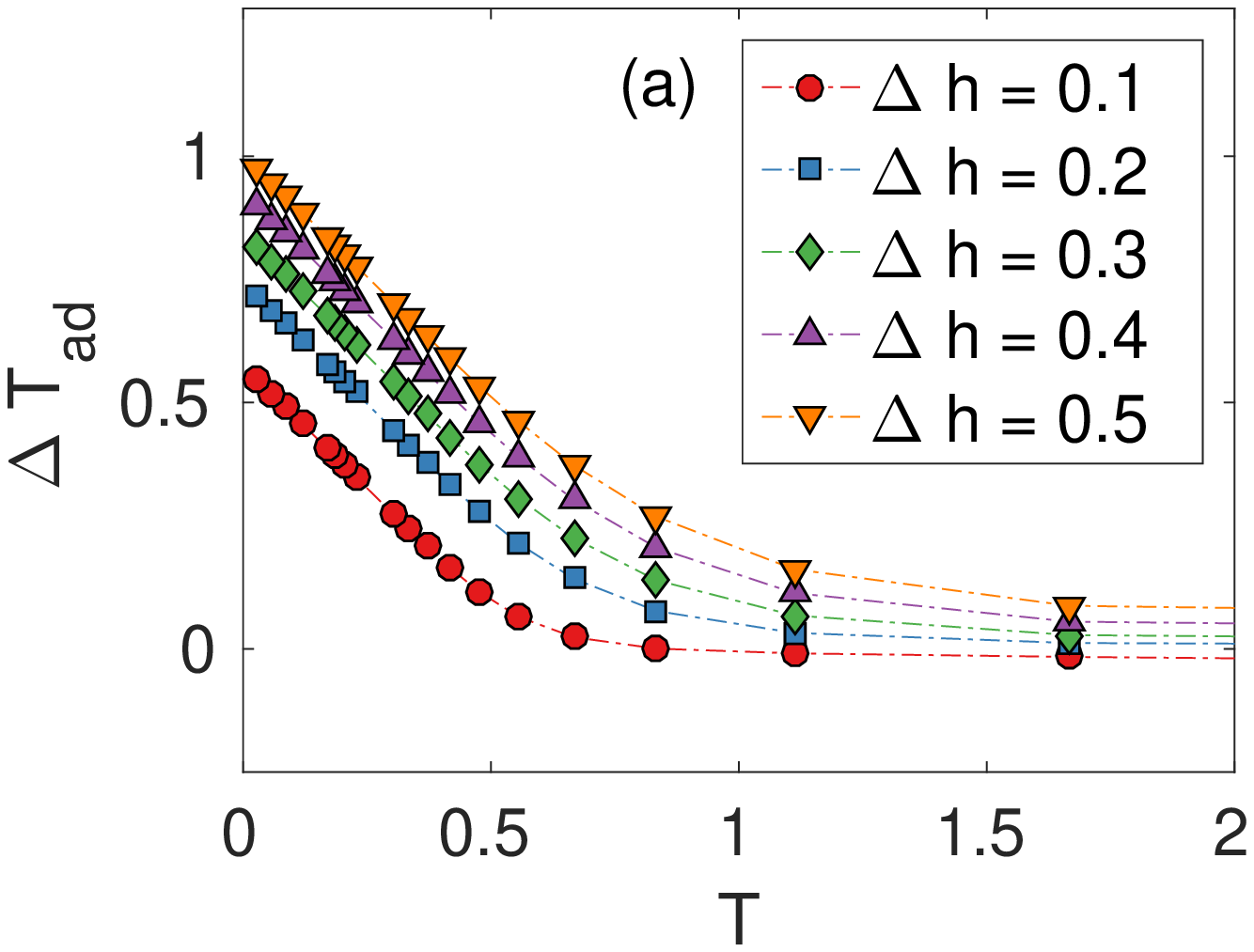}
        %\caption{}
        \label{fig:dt1}
    \end{subfigure}
    \hfill
    \begin{subfigure}[b]{\mywidth}
    	\center
        \centering
        \includegraphics[height = \myheight, clip]{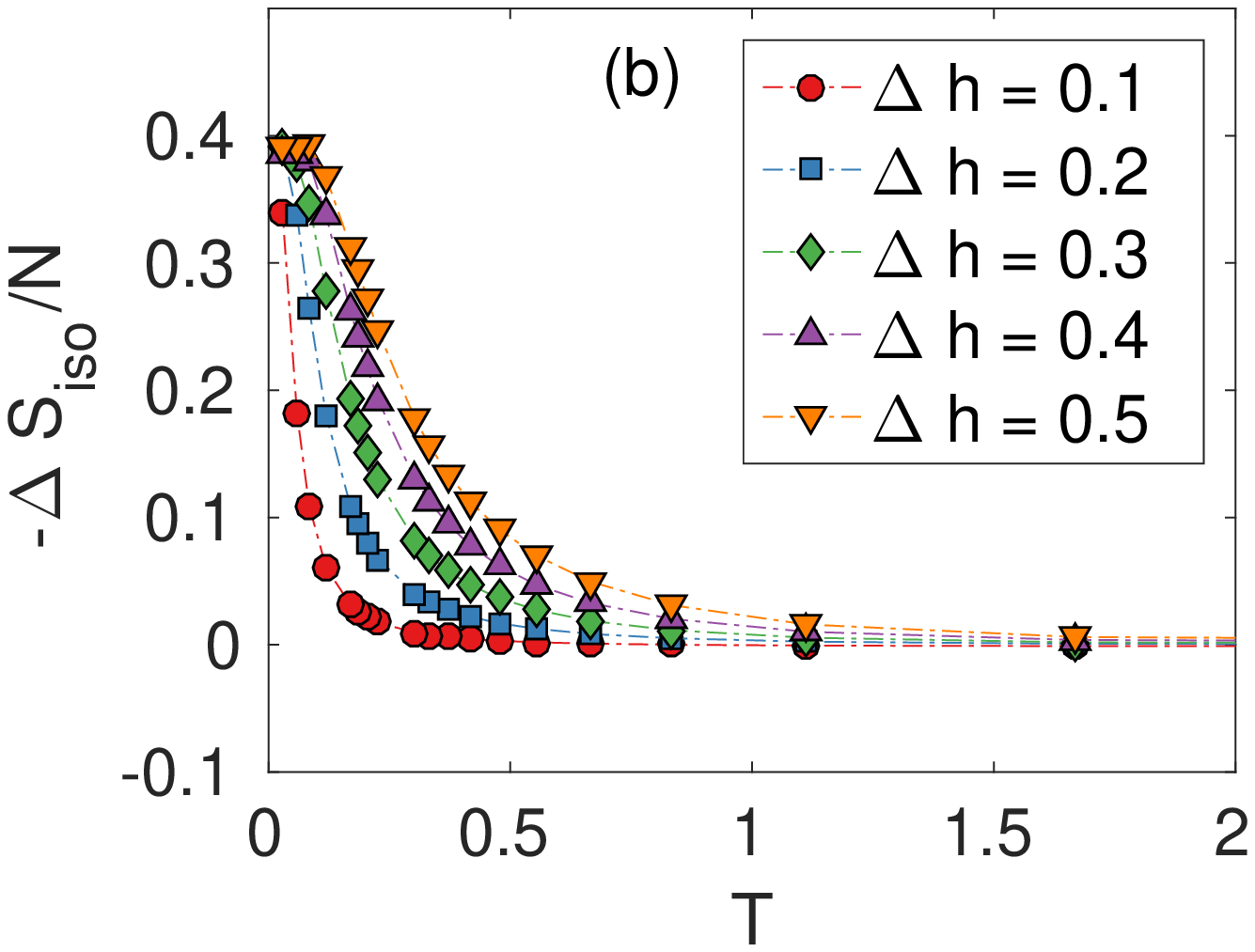}
        %\caption{}
        \label{fig:ds1}
    \end{subfigure}
    \\
    \begin{subfigure}[b]{\mywidth}
    	\center
        \centering
        \includegraphics[height = \myheight, clip]{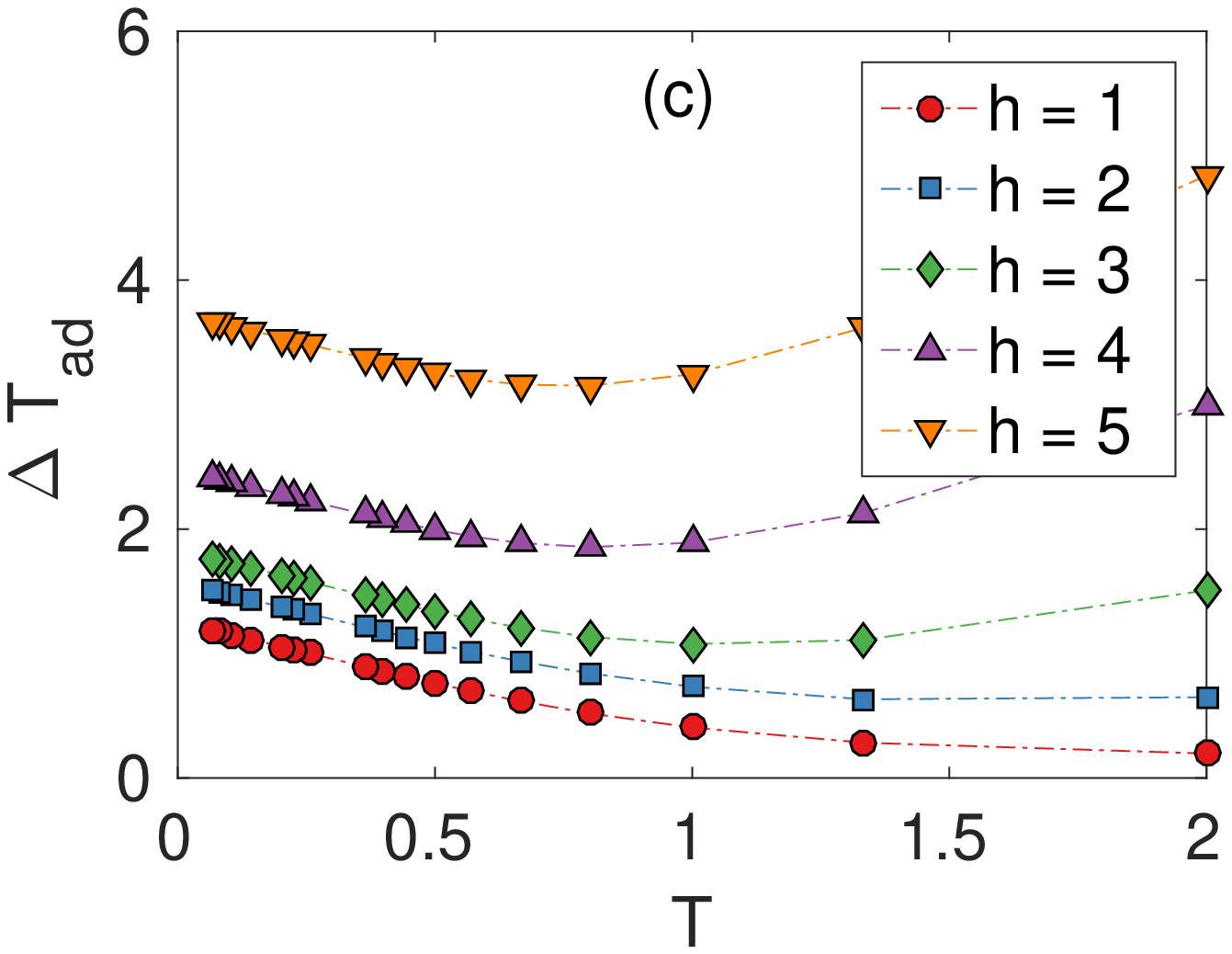}
        %\caption{}
        \label{fig:dt2}
    \end{subfigure}
    \hfill
    \begin{subfigure}[b]{\mywidth}
    	\center
        \centering
        \includegraphics[height = \myheight, clip]{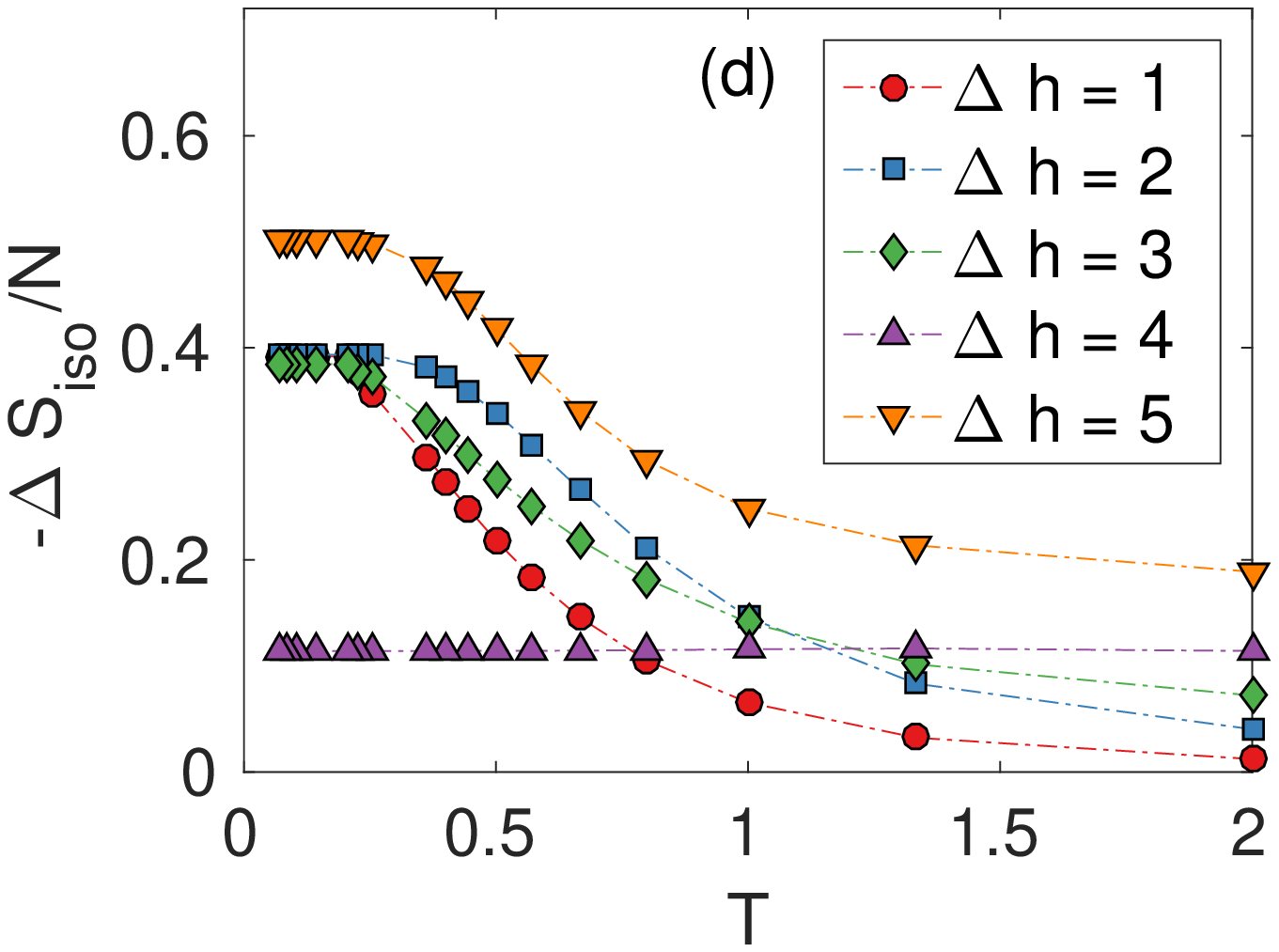}
        %\caption{}
        \label{fig:ds2}
    \end{subfigure}
    \caption{Adiabatic temperature change and isothermal entropy change for various values of the field change $\Delta h$.}
    \label{fig:MCEpot}
\end{figure}

\paragraph{}
In the zero field, the system being antiferromagnetic, has zero magnetization (Fig. \ref{fig:mag} (a)). In the presence of a small field, we observe in GS 1/3 plateau which persists up to $h = 4$. Right at the $h = 4$ the magnetization jumps to the value $m = 3/5$ (Fig. \ref{fig:mag} (b)) and for $h > 4$ , the system reaches the fully saturated state with $m = 1$. This behavior is also reflected in the entropy density (see Figs. \ref{fig:mag} (c) and \ref{fig:mag} (d)), which is in the zero field equal to the theoretical value $0.5018$ \cite{kano1953}. Small fields partially lift the degeneracy and the entropy density reaches the value $0.109$ (which is close to the value mentioned in Ref. \cite{loh2008}). At the sturation field $h = 4$, the entropy density is equal to $0.3878$ and for $h > 4$ it becomes zero. 
\iffalse
Specific heat exibits two peaks. Low-temperature peak moves toward higer temperatures with increasing field $h$. Another high-temperature peak is present even without magnetic field around $T\approx 2$ and it is connected to the short-range order. With increasing $h$ peaks merge.
\fi
\paragraph{}
In addition to the previous quantites, the isothermal entropy change $\Delta S_{iso}$ and the adiabatic temperature change $\Delta T_{ad}$ were calculated from Eqs. (\ref{eq:ds}) and (\ref{eq:dt}), respectively. We chose $h_1 = 0$ and $h_2 > 0$. From Fig. \ref{fig:MCEpot}  one can see that MCE is the most prominent in the region of low temperatures and low fields. Even a small change of the field can lead to a large change of $\Delta S_{iso}$ and $\Delta T_{ad}$. The entropy change has the largest value $-\Delta S_{iso}^{max}/N = 0.3911$ for small fields ($h < 4$), for $h = 4$ it is $-\Delta S_{iso}^{max}/N = 0.1142$ and for $h > 4$ its value is  $-\Delta S_{iso}^{max}/N = 0.5019$. This suggests that IAKL could be used for magnetic refrigeration in a low-temperature region. Similarly, the adiabatic temperature change $\Delta T_{ad}$ has the most interesting behavior in the low-temperature region (Figs. \ref{fig:MCEpot} (a) and (c)). If the field is increased (decreased), the temperature of the system under adiabatic conditions increases (decreases) proportionally to the field's strength.

\section*{Acknowledgment}
This work was supported by the Scientific Grant Agency of Ministry of Education of Slovak Republic (Grant No. 1/0531/19) and Slovak Research and Development Agency (Grant No. APVV-16-0186).

%\end{multicols}

%\printbibliography


\begin{thebibliography}{9}
    \bibitem{diep2013}
    H. T. Diep, \emph{Frustrated Spin Systems}, DOI: 10.1142/8676 World Sci. Pub. Co., Singapore 2013.

    \bibitem{zhitomirsky2003}
    M. E. Zhitomirsky, \emph{Phys. Rev. B} \textbf{67}, 104421 (2003). DOI: 10.1103/PhysRevB.67.104421
        
    \bibitem{syozi1951}
    I. Syozi, \emph{Prog. Theor. Phys.}  \textbf{6} 306 (1951). DOI: 10.1143/ptp/6.3.306
        
    \bibitem{kano1953}
    K. Kano, S. Naya, \emph{Progr. Theor. Phys.} \textbf{10} 158 (1953). DOI: 10.1143/ptp/10.2.158
        
    \bibitem{shores2005}
    M. P. Shores, E. A. Nytko, B. M. Bartlett, D. G. Nocera, \emph{J. Am. Chem. Soc.} \textbf{127} 13462 (2005). DOI: 10.1021/ja053891p
        
    \bibitem{wannier1950}
    G. H. Wannier, \emph{Phys. Rev.}
    \textbf{79} 357 (1950). DOI: 10.1103/PhysRev.79.357
    
    \bibitem{kirkpatrick1977}
    S. Kirkpatrick, \emph{Phys. Rev. B} \textbf{16} 4630 (1977). DOI: 10.1103/PhysRevB.16.4630
    
    \bibitem{loh2008}
    Y. L. Loh, D. X. Yao, W. E. Carlson, \emph{Phys. Rev. B}
    \textbf{77} 134402 (2008). DOI: 10.1103/PhysRevB.77.134402
\end{thebibliography}
\end{document}